\documentclass[aps,pre,amsmath,lengthcheck,superscriptaddress]{revtex4-2}

\usepackage{graphicx}\graphicspath{ {figures/} }
\usepackage{hyperref}
\hypersetup{colorlinks,allcolors=blue,breaklinks}

\newcommand{\be}{\begin{equation}}
\newcommand{\ee}{\end{equation}}
\newcommand{\ba}{\begin{align}}
\newcommand{\ea}{\end{align}}
\newcommand{\bi}{\begin{itemize}}
\newcommand{\ei}{\end{itemize}}

\newcommand{\la}{\left\langle}
\newcommand{\ra}{\right\rangle}
\newcommand{\pd}{\partial}

\newcommand{\bla}{bla\\bla\\bla\\bla\\bla}

\newcommand{\mb}[1]{\mbox{\boldmath$#1$}}
\newcommand{\mc}[1]{\mathcal{#1}}

\begin{document}

\title{Optimal work fluctuations for thermally isolated systems in weak processes}

\author{Pierre Naz\'e}
\email{pierre.naze@unesp.br}

\affiliation{\it Universidade Estadual Paulista, 14800-090, Araraquara, S\~ao Paulo, Brazil}

\date{\today}

\begin{abstract}

The fluctuation-dissipation relation for the classical definition of work is extended to thermally isolated systems, in classical and quantum realms. From this, the optimal work variance is calculated, showing it achieves its minimum possible value, independent of the rate of the process, in a so-called quasistatic variance, related to the difference between the quasistatic work and the difference of Helmholtz's free energy of the system. The result is corroborated by the example of the classical and driven harmonic oscillator, whose probability density function of the work distribution is non-Gaussian and constant for different rates. The optimal variance is calculated for the quantum Ising chain as well, showing its finiteness if the linear response validity criterion is complied. A stronger definition of the arbitrary constant for the relaxation function of thermally isolated systems is obtained along the work.

\end{abstract}

\maketitle

\section{Introduction}
\label{sec:intro}

Fluctuation-dissipation relations are important identities able to furnish information about optimal control of dissipated averages and fluctuations. In the context of classical and isothermal processes, it has been shown for quadratic potentials~\cite{jarzynski1997}, slowly-varying~\cite{speck2004distribution} and weak~\cite{naze2023optimal} processes. Using the quantum definition of the work, it has been shown its breakdown for slowly-varying and weak processes~\cite{miller2019work,guarnieri2023generalised}.

The aim of this work is to obtain the optimal work fluctuations of classical and quantum thermally isolated systems using the classical definition of work. This is done by means of an extension of the fluctuation-dissipation relation. By contrast with the isothermal case, such a relation presents a breakdown, presenting an extra quasistatic variance in the equality, which is independent of the rate of the process, and related to the difference between the quasistatic work and difference of Helmholtz's free energy of the system. When the protocol is optimal, the optimal variance achieves its minimal value in this quasistatic variance. To exemplify it, it is presented the driven harmonic oscillator. In particular, its work probability distribution is non-Gaussian and independent of the rate. The optimal variance for the quantum Ising chain is calculated as well, showing its finiteness if the the linear response validity agreement is complied.

\section{Weak processes}
\label{sec:lrt}

I start defining notations and developing the main concepts to be used in this work. This section is based on the technical introductory section of Ref.~\cite{naze2023optimal}.

Consider a classical system with a Hamiltonian $\mc{H}(\mb{z}(\mb{z_0},t)),\lambda(t))$, where $\mb{z}(\mb{z_0},t)$ is a point in the phase space $\Gamma$ evolved from the initial point $\mb{z_0}$ until time $t$, with $\lambda(t)$ being a time-dependent external parameter. Initially, the system is at equilibrium with a heat bath of temperature $\beta\equiv {(k_B T)}^{-1}$, where $k_B$ is Boltzmann's constant. The heat bath is then removed from the system, and during a switching time $\tau$, the external parameter is changed from $\lambda_0$ to $\lambda_0+\delta\lambda$. The average work performed on the system during this interval of time is
\be
\overline{W}(\tau) \equiv \int_0^\tau \la\pd_{\lambda}\mc{H}(t)\ra_0\dot{\lambda}(t)dt,
\label{eq:work}
\ee
where $\partial_\lambda$ is the partial derivative in respect to $\lambda$ and the superscripted dot the total time derivative. The generalized force $\la\pd_{\lambda}\mc{H}\ra_0$ is calculated using the averaging $\langle\cdot\rangle_0$ over the initial canonical ensemble. The external parameter can be expressed as
\be
\lambda(t) = \lambda_0+g(t)\delta\lambda,
\label{eq:ExternalParameter}
\ee
where to satisfy the initial conditions of the external parameter the protocol $g(t)$ must satisfy the following boundary conditions $g(0)=0$, $g(\tau)=1$.

Linear-response theory aims to express average quantities until the first-order of some perturbation parameter considering how this perturbation affects the observable to be averaged and the probabilistic distribution \cite{kubo2012}. In our case, we consider that the parameter does not considerably change during the process, $|g(t)\delta\lambda/\lambda_0|\ll 1$, for all $t\in[0,\tau]$ and $\lambda_0\neq 0$. The generalized force can be approximated until the first order as~\cite{naze2020}
\begin{equation}
\begin{split}
\la\pd_{\lambda}\mc{H}(t)\ra_0 =&\, \la\pd_{\lambda}\mc{H}\ra_0-\widetilde{\Psi}_0 \lambda(t)\\
&+\int_0^t \Psi_0(t-t')\dot{\lambda}(t')dt',
\label{eq:genforce-relax}
\end{split}
\end{equation}
where 
\be
\Psi_0(t) = \beta\la\pd_\lambda\mc{H}(0)\pd_\lambda\mc{H}(t)\ra_0-\mc{C}
\ee 
is the relaxation function and $\widetilde{\Psi}_0\equiv \Psi_0(0)-\la\pd_{\lambda\lambda}^2\mc{H}\ra_0$ \cite{kubo2012}. The constant $\mc{C}$ is arbitrary, whose chosen value I am going to discuss in the next section. Combining Eqs.~\eqref{eq:work} and \eqref{eq:genforce-relax}, the average work performed at the linear response of the generalized force is
\begin{equation}
\begin{split}
\overline{W}(\tau) = &\, \delta\lambda\la\pd_{\lambda}\mc{H}\ra_0-\frac{\delta\lambda^2}{2}\widetilde{\Psi}_0\\
&+\frac{1}{2}\int_0^\tau\int_0^\tau \Psi_0(t-t')\dot{\lambda}(t')\dot{\lambda}(t)dt'dt.
\label{eq:work2}
\end{split}
\end{equation}
where the symmetric property of the relaxation function was used \cite{kubo2012}. Such an equation holds for finite-time and weak processes. 

Our treatment throughout this work will be classical, but the same reasoning with similar arguments leads to the same average work for quantum systems, where the observables become operators, and averages, traces.

\section{Constant $\mathcal{C}$}

In previous works, I have observed that the double integral on Eq.~\eqref{eq:work2} depends on the path, which would indicate that the other terms are the contribution of the quasistatic work $W_{\rm qs}$. However, the constant $\mathcal{C}$ must be chosen properly to produce such a result. For isothermal processes, it is chosen such that the relaxation function decorrelates for long times
\be
\lim_{t\rightarrow \infty}\Psi_0(t)=0,
\ee
which is nothing more than a feature of the Second Law of Thermodynamics. However, for thermally isolated systems, such operation does not make any sense, because the relaxation function does not decorrelate. One alternative is the definition proposed by Kubo~\cite{kubo2012} where $\mathcal{C}$ is calculated such that
\be
\lim_{s\rightarrow 0^+}s\widetilde{\Psi}_0(s)=0,
\ee
where $\widetilde{\cdot}$ is the Laplace transform. This definition, in my opinion, although the success verified {\it a posteriori}~\cite{acconcia2015,bonancca2016non,myers2021,naze2022kibble}, lacks an {\it a priori} physical motivation. In what follows I propose an alternative which will furnish a value to $\mathcal{C}$ agreeing with the Second Law of Thermodynamics.

\section{Cumulant series}

Jarzynski's equality is well recognized as a generalization of the Second Law of Thermodynamics~\cite{jarzynski1997}. I am going to propose a definition of $\mathcal{C}$ which will agree with such a relation. According to it, it holds the following cumulant series expansion for the irreversible work
\be
\beta W^{\rm irr}=\beta(\overline{W}-\Delta F)=\sum_{n=2}^{\infty}\frac{(-\beta)^n}{n!}\kappa_n,
\ee
where $\overline{W}$ is the average work for a thermally isolated system, $\kappa_n$ the cumulants for the work probability distribution and $\Delta F$ is the difference of Helmholtz's free energy. Writing in terms of the excess work, one has
\begin{align}
\beta W^{\rm ex}&=\beta(\overline{W}-W_{\rm qs})\\
&=\sum_{n=2}^{\infty}\frac{(-\beta)^n}{n!}\kappa_n+\beta(\Delta F-W_{\rm qs})    
\end{align}
In particular, using linear response theory, one has
\be
\beta W^{\rm ex}_2=\frac{\beta^2}{2}\kappa_2+\beta(\Delta F-W_{\rm qs})_2,
\ee
where the terms were calculated until the second order in the parameter perturbation. Using 
\be
\beta W^{\rm ex}_2-\frac{\beta^2}{2}\kappa_2=-\beta\mathcal{C}-\frac{\beta^2}{2}\delta\lambda^2\langle \partial_\lambda \mathcal{H}(0)\rangle_0^2
\ee
one has
\be
\beta\mathcal{C}=-\frac{\beta^2}{2}\delta\lambda^2\langle \partial_\lambda \mathcal{H}(0)\rangle_0^2+\beta(W_{\rm qs}-\Delta F)_2.
\ee
However, by the definitions of $\Delta F$ and $W_{\rm qs}$, one has
\be
\beta(W_{\rm qs}-\Delta F)_2=\frac{\beta^2}{2}\delta\lambda^2\langle \partial_\lambda \mathcal{H}(0)\rangle_0^2.
\ee
Therefore
\be
\mathcal{C}=0,
\ee
which is a stronger and physically more meaningful definition for such a constant than that proposed by Kubo. Such a result is corroborated by different works, for classical and quantum systems~\cite{acconcia2015,bonancca2016non,myers2021,naze2022kibble}. Using the classical definition of work, the cumulant series can be extended using the quantum treatment. 

\section{Fluctuation-dissipation relation}

From the approximation of the cumulant series for linear response theory deduced in the previous section, observe that it holds the following fluctuation-dissipation relation
\be
\beta W^{\rm ex}_2=\frac{\beta^2}{2}\sigma^2_{W_2}-\frac{\beta^2\delta\lambda^2}{2}\langle \partial_\lambda \mathcal{H}(0)\rangle_0^2,
\label{eq:FDR}
\ee
where $\sigma^2_{W_2}$ is the variance of the work probability distribution calculated until the second-order in the parameter perturbation. That relation implies that the excess work expends less energy than its irreversible work counterpart. The breakdown in the relation when compared to isothermal cases occurs due to the difference between the quasistatic work and the difference of Helmholtz's free energy.

\begin{figure}[t]
    \centering
    \includegraphics[scale=0.45]{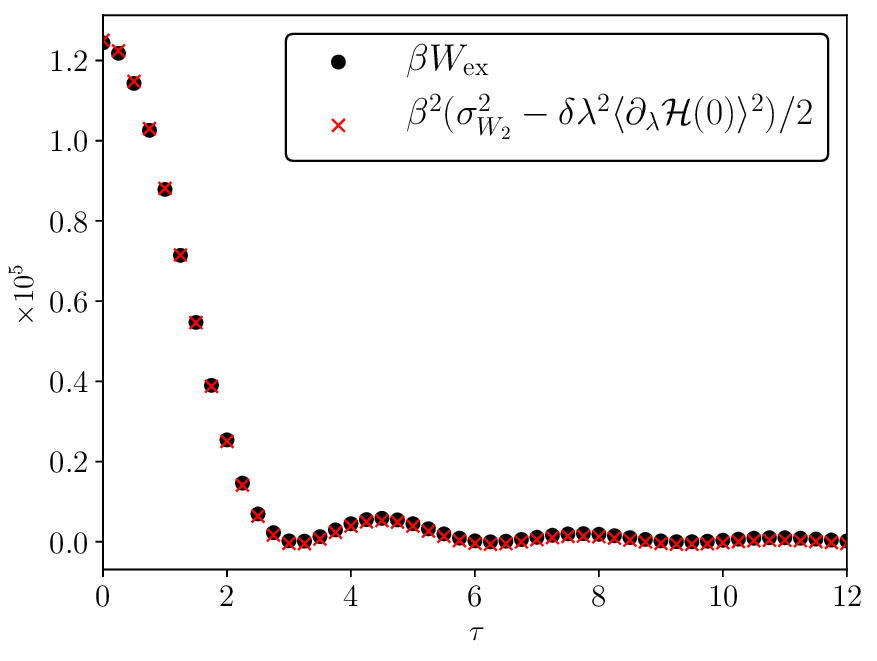}
    \caption{Fluctuation-dissipation relation for the linear driven harmonic oscillator. It was used $\delta\lambda/\lambda_0=0.01$. Here, $\delta\lambda^2\langle\partial_\lambda \mathcal{H}(0)\rangle_0^2=2.5\times 10^{-5}$.}
    \label{fig:1}
\end{figure}

To exemplify such a result, consider a linear-driven harmonic oscillator, whose Hamiltonian is
\be
\mathcal{H}(\lambda(t))=\frac{p^2}{2}+\lambda(t)\frac{q^2}{2},
\ee
where $\lambda(t)=\lambda_0+\delta\lambda(t/\tau)$. Its solution is known for the full dynamics, from where the average work and work variance can be calculated. Also, the quasistatic work is known~\cite{acconcia2015}
\be
W_{\rm qs}=\frac{1}{\beta}\left(\sqrt{\frac{\lambda_0+\delta\lambda}{\lambda_0}}-1\right).
\ee
Considering $\delta\lambda/\lambda_0=0.01$, Fig.~\ref{fig:1} depicts the fluctuation-dissipation relation expressed in Eq.~\eqref{eq:FDR}. Here, $\delta\lambda^2\langle\partial_\lambda \mathcal{H}(0)\rangle_0^2=2.5\times 10^{-5}$.

\section{Optimal work fluctuations}

It has been shown that for thermally isolated performing weak processes the optimal excess work is null~\cite{naze2023universal}. Therefore, by the fluctuation-dissipation relation, the optimal variance of the work for thermally isolated systems achieves its minimum possible value, given by
\be
\sigma_{W_2}^{2*}=\delta\lambda^2\langle \partial_\lambda \mathcal{H}(0)\rangle_0^2,
\ee
which is independent of the rate of the process. It is indeed an intrinsic characteristic of the system. One may consider it as a quasistatic variance for the thermally isolated system. This unexpected result shows that although one achieves by the optimal protocol the quasistatic work for arbitrary rates, there is always an intrinsic error associated. In particular, it is expected that $\sigma_{W_2}^{2*}(\tau)\propto 1/\beta^2$, because of the average on the canonical ensemble. In this situation, if the system starts at $T\approx 0$, the variance diverges.

\begin{figure}[h]
    \centering
    \includegraphics[scale=0.45]{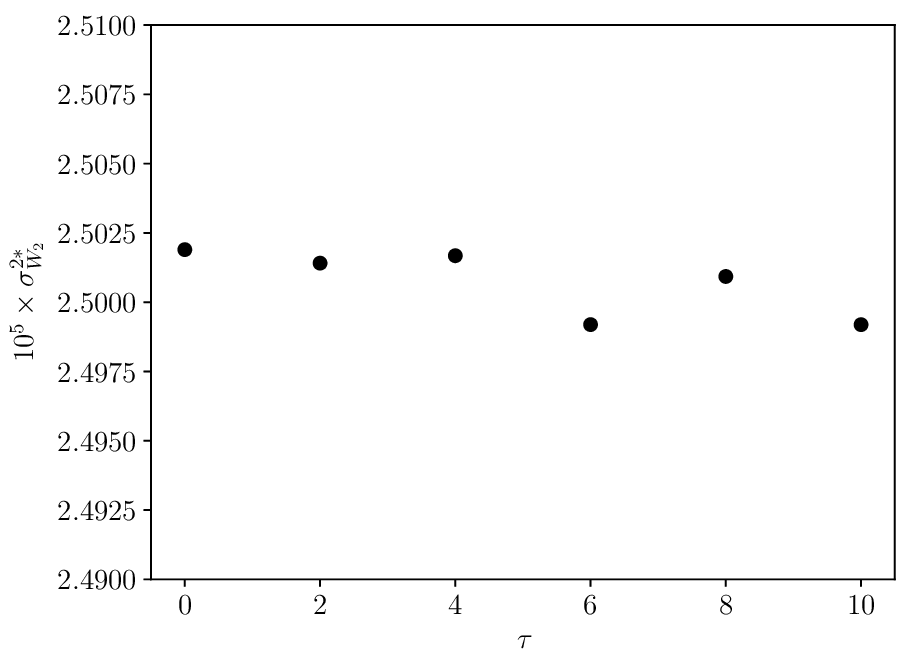}
    \caption{Optimal work variance for driven harmonic oscillator. Here, $\beta^2\delta\lambda^2\langle\partial_\lambda \mathcal{H}(0)\rangle_0^2=2.5\times 10^{-5}$. It was used $10^5$ initial conditions and $\delta\lambda/\lambda_0=0.01$.}
    \label{fig:2}
\end{figure}

\begin{figure}[h]
    \centering
    \includegraphics[scale=0.45]{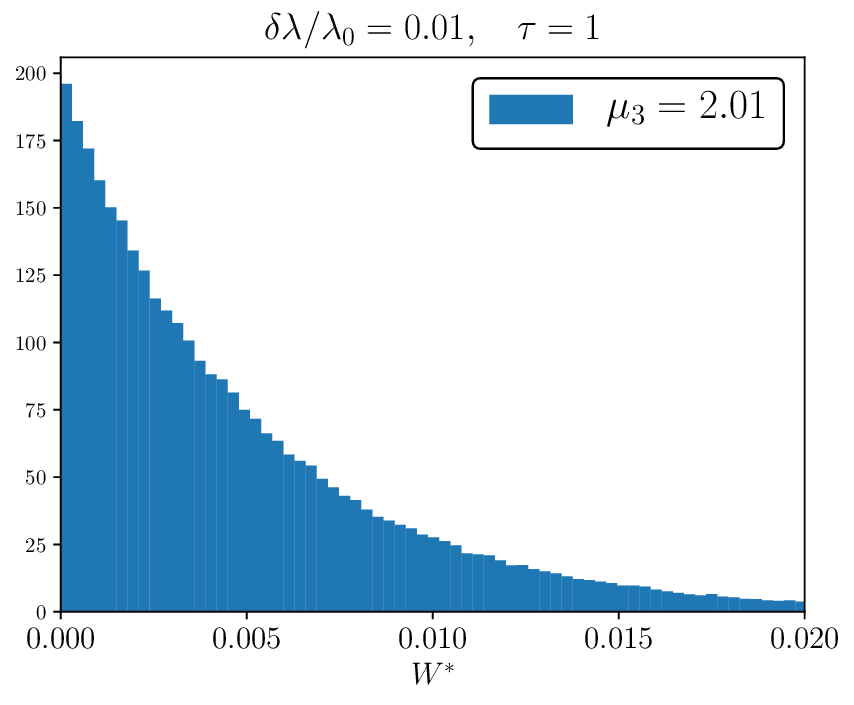}
    \caption{Optimal work probability density function for driven harmonic oscillator. Here, $\delta\lambda/\lambda_0$=0.01 for $10^5$ initial conditions. The histogram is non-Gaussian ($\mu_3>0$) and does not change with different rates.}
    \label{fig:3}
\end{figure}

To exemplify it, consider again the driven harmonic oscillator, but driven with the optimal protocol for linear response, given by~\cite{acconcia2015,naze2023universal}
\be
g^*(t)=\frac{t}{\tau}+\frac{\delta(t)-\delta(\tau-t)}{4\lambda_0\tau}.
\ee
The optimal work variance is exhibited in Fig.~\ref{fig:2} for different rates. In this particular case, $\beta^2\delta\lambda^2\langle\partial_\lambda \mathcal{H}(0)\rangle_0^2=2.5\times 10^{-5}$. Figure~\ref{fig:3} depicts the optimal probability distribution function of the work, which is also non-Gaussian~\cite{naze2023optimal} and constant for different rates. I used in the computational simulation $10^5$ initial conditions, and $\delta\lambda/\lambda_0=0.01$.

\section{Quantum Ising chain}

One important problem to deal with nowadays is the performance of quantum annealing, aiming to apply in quantum computing~\cite{deffner2019quantum}. In particular, a question not so explored is the work fluctuations present in driving processes. Consider then the quantum Ising model, whose Hamiltonian operator is
\be
\mathcal{H} = -J\sum_{i=1}^{N} \sigma_i^x\sigma_{i+1}^x-\Gamma\sum_{i=1}^{N} \sigma_i^z.  
\label{eq:qim}
\ee
where each one of the $N$ spins has a vector $\vec{\sigma}_i := \sigma_i^x {\bf x}+ \sigma_i^y {\bf y}+ \sigma_i^z {\bf z} $ composed by the Pauli matrices. The parameter $J$ is the coupling energy and $\Gamma$ is the transverse magnetic field. Also, the system is subjected to periodic boundary conditions to an even number of spins. In Ref.~\cite{naze2023universal} I have found the optimal protocol that produces a null excess work of such a system. Under those circumstances, the work fluctuations will be given by the quasistatic variance of the system. In particular, this quantity is
\be
\sigma_{W_2}^{2*}=2\delta \Gamma^2\sum_{n,m=1}^{N/2}\tanh{\beta\epsilon_n}\tanh{\beta\epsilon_m},
\ee
where
\be
\epsilon_n=2\sqrt{J^2+\Gamma^2-2\Gamma J \cos{(\pi(2n-1)/N)}}.
\ee
In particular, for purposes of performance of quantum annealing, it is interesting to observe how the quasistatic variance behaves for a system that starts with $T=0$. In this case, one has
\be
\sigma_{W_2}^{2*}\propto\delta \Gamma^2 N^2(N+1)^2
\ee
which indicates that the quasistatic variance diverges if the system is in the thermodynamic limit, where $N\gg 1$. However, in this situation, linear response only works for very small perturbations~\cite{naze2022kibble}. Therefore, after choosing $\delta\Gamma\propto N^{-2}$, this will compensate the divergence produced by the thermodynamic limit, generating a finite quasistatic variance for the system. Knowing how the quasistatic variance of a system behaves could be an additional criterion to evaluate the validity of linear response in quantum phase transition situations than only using the criterion proposed in Ref.~\cite{naze2022kibble}.

\section{Final remarks}
\label{sec:final}

In this work, in order to find the optimal work fluctuation for thermally isolated systems performing weak adiabatic processes, in classical and quantum realms, for the classical definition of work, the fluctuation-dissipation relation was extended. The equality presents a breakdown in comparison to the isothermal case, where an extra quasistatic variance appears, related to the difference between the quasistatic work and the difference of Helmholtz's free energy of the system. From this, the optimal work variance was calculated showing it achieves its minimum value, independent of the rate of the process. The result was corroborated by the example of the driven harmonic oscillator. The optimal variance for the quantum Ising chain is calculated as well, showing its finiteness if the the linear response validity agreement is complied. The arbitrary constant for the relaxation function of thermally isolated systems was shown to be equal to zero to agree with the Second Law of Thermodynamics.

\bibliography{VTILR.bib}
\bibliographystyle{apsrev4-2}

\end{document}